\documentclass[conference,10pt]{IEEEtran}
\ifCLASSINFOpdf
\else
\fi
\ifCLASSINFOpdf
\else
\fi

\usepackage{array,booktabs}
\usepackage{algpseudocode}
\usepackage{algorithm}
\usepackage{amsfonts}
\usepackage{amsmath}
\usepackage{amssymb}
\usepackage{amsthm}
\usepackage{graphicx}
\usepackage[noadjust]{cite}
\usepackage{mathrsfs}
\usepackage{stmaryrd}
\usepackage{siunitx}
\usepackage{multicol}
\usepackage{tikz}
\usepackage{setspace}
\usepackage{pstricks, pst-node, pst-plot, pst-circ}
\usepackage{moredefs}
\usetikzlibrary{shapes}

\makeatletter
\newtheoremstyle{newcorollary}
  {3pt}
  {3pt}
  {}
  {1em}
  {\itshape}
  {:}
  {.5em}
  {\thmname{#1}\thmnumber{\@ifnotempty{#1}{ }#2}%
   \thmnote{ {\the\thm@notefont(\itshape#3)}}}
\makeatother
\theoremstyle{newcorollary}

\makeatletter
\newtheoremstyle{newdefinition}
  {3pt}
  {3pt}
  {}
  {1em}
  {\itshape}
  {:}
  {.5em}
  {\thmname{#1}\thmnumber{\@ifnotempty{#1}{ }#2}%
   \thmnote{ {\the\thm@notefont(\itshape#3)}}}
\makeatother
\theoremstyle{newdefinition}
\newtheorem*{definition}{Definition}

\makeatletter
\newtheoremstyle{newlemma}
  {3pt}
  {3pt}
  {}
  {1em}
  {\itshape}
  {:}
  {.5em}
  {\thmname{#1}\thmnumber{\@ifnotempty{#1}{ }#2}%
   \thmnote{ {\the\thm@notefont(\itshape#3)}}}
\makeatother
\theoremstyle{newlemma}
\newtheorem{lemma}{Lemma}

\makeatletter
\newtheoremstyle{newtheorem}
  {3pt}
  {3pt}
  {}
  {1em}
  {\itshape}
  {:}
  {.5em}
  {\thmname{#1}\thmnumber{\@ifnotempty{#1}{ }#2}%
   \thmnote{ {\the\thm@notefont(\itshape#3)}}}
\makeatother
\theoremstyle{newtheorem}
\newtheorem{theorem}{Theorem}

\makeatletter
\newtheoremstyle{newproposition}
  {3pt}
  {3pt}
  {}
  {1em}
  {\itshape}
  {:}
  {.5em}
  {\thmname{#1}\thmnumber{\@ifnotempty{#1}{ }#2}%
   \thmnote{ {\the\thm@notefont(\itshape#3)}}}
\makeatother
\theoremstyle{newproposition}

\makeatletter
\newtheoremstyle{newproof}
  {3pt}
  {3pt}
  {}
  {2em}
  {\itshape}
  {:}
  {.5em}
  {\thmname{#1}\thmnumber{\@ifnotempty{#1}{ }#2}%
   \thmnote{ {\the\thm@notefont(\itshape#3)}}}
\makeatother
\theoremstyle{newproof}
\newtheorem*{proofs}{Proof}

\makeatletter
\newtheoremstyle{newremark}
  {3pt}
  {3pt}
  {}
  {2em}
  {\itshape}
  {:}
  {.5em}
  {\thmname{#1}\thmnumber{\@ifnotempty{#1}{ }#2}%
   \thmnote{ {\the\thm@notefont(\itshape#3)}}}
\makeatother
\theoremstyle{newproof}

\providelength{\AxesLineWidth}       
\providelength{\plotwidth}           
\providelength{\LineWidth}           
\providelength{\MarkerSize}          
\newrgbcolor{GridColor}{0.8 0.8 0.8}%

\begin{document}

\title{Rate Allocation for Decentralized Detection\\in Wireless Sensor Networks}

\author{
\IEEEauthorblockN{Alla Tarighati, and Joakim Jald{\'e}n}
\IEEEauthorblockA{ACCESS Linnaeus Centre, Department of Signal Processing, \\
KTH Royal Institute of Technology, Stockholm, Sweden\\
Email: \{allat, jalden\}@kth.se}
}

\maketitle

\begin{abstract}
We consider the problem of decentralized detection where peripheral nodes make noisy observations of a phenomenon and send quantized information about the phenomenon towards a fusion center over a sum-rate constrained multiple access channel. The fusion center then makes a decision about the state of the phenomenon based on the aggregate received data. Using the Chernoff information as a performance metric, Chamberland and Veeravalli previously studied the structure of optimal rate allocation strategies for this scenario under the assumption of an unlimited number of sensors. Our key contribution is to extend these result to the case where there is a constraint on the maximum number of active sensors. In particular, we find sufficient conditions under which the uniform rate allocation is an optimal strategy, and then numerically verify that these conditions are satisfied for some relevant sensor design rules under a Gaussian observation model.
\end{abstract}

\IEEEpeerreviewmaketitle
\section{Introduction}\label{sec:intro}
Decentralized detection is a central problem in wireless sensor networks (WSN) \cite{Veer12,Cham03,Cham07,li07}. In a decentralized detection problem, spatially separated sensors make private noisy observations of the state of a phenomenon and send their observations to a fusion center (FC) over rate constrained channels for the final decision about the state of the nature. 
This problem has been considered extensively in the literature when each sensor has a private communication link to transmit its information towards the FC, see \cite{Veer12} and references therein.

However, in a wireless sensor network the sensors typically share a common multiple access channel (MAC) to the FC. In this work we assume that the MAC channel is error-free but subject to a common sum-rate constraint of rate $R$ bits per channel use. We consider a binary hypothesis testing problem under which $N$ sensors make private observations of the phenomenon, or hypothesis, $H \in \mathcal{H}\triangleq\{H_0,H_1\}$. 
Conditioned on the true hypothesis $H$, the observations at the sensors are independent and identically distributed (iid). Each sensor $S_n$, for $n=1,\ldots,N$, is required to quantize its own observation into an $r_n$ (integer) bit message in such a way that the sum-rate constraint
\begin{equation}
\sum_{n=1}^{N}r_n \leq R
\label{eq:Rconstraint}\end{equation}
is satisfied.
The FC then uses the aggregate set of the received messages to make the final decision $\hat{H}\in\mathcal{H}$. 

Chamberland and Veeravalli \cite{Cham03} studied this network model
in terms of the optimal number of sensors $N$ and rate allocation $\underline{r} = (r_1,\ldots,r_N)$ using the Chernoff information at the input of the FC as a performance metric. They proved that if, for a given observation model, there exists a rate-one quantization rule for a single sensor which leads to the transfer of at least half of the Chernoff information contained in each raw observation, then having $N=R$ rate-one sensors is optimal. They also proved that such a rate-one sensor decision rule exists when the observations at the sensors are equal variance Gaussian or Exponentially distributed.

Although the optimality result of \cite{Cham03}  greatly simplifies the network design it may in practice, due to cost and space constraints, not always be feasible to have $N=R$ sensors in a network. In the present paper we therefore address the problem of finding an optimal rate allocation $\underline{r}$ when
the total number of sensors $N$ is fixed a priori. For simplicity, we assume that $R=MN$, where $M$ in a positive integer. As in \cite{Cham03}, we use the Chernoff information at the input of the FC as the performance metric. We will show that if for a given sensor design method, the Chernoff information at the output of a single sensor is a discrete concave function of the sensor's rate, then uniform rate allocation is an optimal strategy for the network. We will also argue and show numerically that existing sensor design rules do in fact yield a per sensor Chernoff information that is a discrete concave function of the rate. We will finally illustrate numerically how this translates into network performance in terms of error probability.

The outline of this paper is as follows. In Section \ref{sec:pre} we formulate the problem. Then, in Section \ref{sec:res} we present our main results and establish the numerical results in Section \ref{sec:sim}.  The paper is concluded in Section \ref{sec:con}.

\section{Preliminaries}\label{sec:pre}

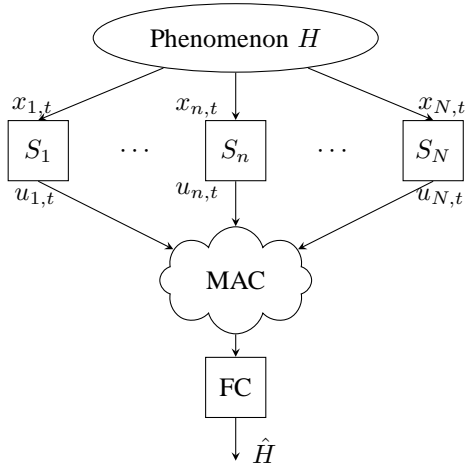
\begin{figure}[t]
\centering
\begin{tikzpicture}[align=center,scale=0.75,>=stealth] 
\node (S1) at (-3.5,3) [rectangle,minimum size=0.8cm,draw] {$S_1$};
\node (dots1) at (-1.75,3) {$\cdots$};
\node (Sn) at (0,3) [rectangle,minimum size=0.8cm,draw] {$S_n$};
\node (dots2) at (1.75,3) {$\cdots$};
\node (SN) at (3.5,3) [rectangle,minimum size=0.8cm,draw] {$S_N$};
\node (PH) at (0,5) [ellipse,inner sep=2mm,draw] {Phenomenon $H$};
\draw [->] (PH) -- (S1.north) node  [near end,left,inner sep=6pt] {$x_{1,t}$};
\draw [->] (PH) -- (Sn.north) node  [near end,left,inner sep=6pt] {$x_{n,t}$};
\draw [->] (PH) -- (SN.north) node  [near end,right,inner sep=6pt] {$x_{N,t}$};
\node (MAC) at (0,.7) [cloud, draw,cloud puffs=8,cloud puff arc=150, aspect=2, inner ysep=.6em] {MAC};
\draw [->] (S1.south) -- (MAC) node  [near start,left,inner sep=6pt] {$u_{1,t}$};
\draw [->] (Sn.south) -- (MAC) node  [near start,left,inner sep=6pt] {$u_{n,t}$};
\draw [->] (SN.south) -- (MAC) node  [near start,right,inner sep=6pt] {$u_{N,t}$};
\node (FC) at (0,-1.2) [rectangle,minimum size=0.8cm,draw] {FC};
\draw [->] (MAC) -- (FC) node  [near start,right,inner sep=7pt] {};
\draw [->] (FC) -- (0,-2.5) node  [near end,right,inner sep=6pt] {$\hat{H}$};
\end{tikzpicture} 
\caption{Setting of sensors in a network, where the sensors send their data through a MAC channel to the FC.}
\label{fig:topology}
\end{figure}

We consider a binary hypothesis testing problem where $N$ sensors, $S_1,\ldots,S_N$, are arranged as in Fig.~\ref{fig:topology}. Sensor $S_n$ makes at each time $t = 1,\ldots,T$ an observation $x_{n,t}$ about the state of the same phenomenon $H$, and computes a message $u_{n,t}$ for the FC using its decision function $\gamma_{n}$. We assume that observations are iid over space and time, i.e., $x_{n,t}$ is viewed as independent realizations of a common random variable $X$ with conditional probability density function (pdf) $f_{X|H}(x|H_j)$, where $j=0,1$, over some observation space $\mathcal{X}$.
The output message $u_{n,t} = \gamma_n(x_{n,t})$ is from an $r_n$-bit message set $\mathcal{U}_{r_n} \triangleq \{1, 2,\ldots,2^{r_n}\}$, where\footnote{We use $\mathbb{Z}^{+}$ to denote the set of natural numbers excluding 0, i.e., $\mathbb{Z}^{+} \triangleq \{1, 2, \ldots \}$, and let $\mathbb{Z}^{++} \triangleq \mathbb{Z}^{+} \cup \{ 0 \}$.} $r_n \in \mathbb{Z}^{++}$ is the allocated rate to sensor $S_n$. The FC makes the final decision in favor of a hypothesis based on the aggregated set of sensor messages over space and time using its function $\gamma_0:{\underline{\mathcal{U}}}\to \mathcal{H}$, i.e.,
$\gamma_0\left(u_{1,1:T},\ldots,u_{N,1:T} \right)=\hat{H}$,
where $\underline{\mathcal{U}}\triangleq \mathcal{U}_{r_1} \times \ldots \times \mathcal{U}_{r_N}$ and $u_{n,1:T}\triangleq \left(u_{n,1}\ldots,u_{n,T}\right)$. It is well known that the optimal FC rule, in the sense that it minimizes the Bayesian probability of error $P_\mathrm{E}^{(T)}=\Pr\left(\hat{H}\neq H \right)$, is given by the maximum a-posteriori (MAP) detector, which can be implemented as a likelihood ratio test on the aggregate set of sensor messages \cite{Cham03}. Our main focus is on the properties of the optimal rate allocation $\underline{r}$ subject to the sum-rate constraint \eqref{eq:Rconstraint}, under the assumption of a given sensor design rule for each allocated rate and under optimal FC processing.

The output messages of sensor $S_n$ over time can be viewed as realizations of a random variable $U_n$ with a conditional probability mass function (pmf) given by
\begin{equation*}\begin{split}
P_{U_n\vert H}\left(u_n\vert H_j \right)
&=\Pr\big\{x_n:\gamma_{n}\left(x_n\right)=u_n\vert H_j \big\}\\
&=\int_{x\in \gamma^{-1}_n\left(u_n \right)}\! f_{X\vert H}\left(x\vert H_j\right)\,dx\,,
\end{split}\end{equation*}
where $\gamma^{-1}_n\left(u_n \right)$ denotes the set of observations $x$ that satisfy $\gamma_{n}(x)=u_n$. The aggregate set of sensor messages at any particular time $t$ may similarly be viewed as a realization of a random vector $\underline{U}$ with pmf
\begin{equation}\begin{split}
P_{\underline{U}\vert H}\left(\underline{u}\vert H_j\right)
&=P_{U_1\vert H}(u_1\vert H_j)\ldots P_{U_N\vert H}(u_N\vert H_j)\\
&=\prod_{n=1}^{N}\int_{x\in \gamma_{n}^{-1}(u_n)}\!f_{X\vert H}(x\vert H_j)\,dx\, ,
\end{split}\end{equation}
where $\underline{u}\triangleq(u_1,\ldots,u_N)$.
Given a decision function $\gamma_{n}$ for a sensor $S_n$ of rate $r_n$, let
\begin{align}
\mathcal{C}_{r_n}\left(\gamma_{n},\alpha \right) & \triangleq \label{eq:Calpha} \\
-\log &\sum_{u_n\in \mathcal{U}_{r_n}}\left[P_{U\vert H}\left({u_n}\vert H_0\right)\right]^\alpha\left[P_{{U}\vert H}\left({u_n}\vert H_1\right)\right]^{1-\alpha}\,. \nonumber
\end{align}
The \emph{Chernoff information} of sensor $S_n$ is then given by
\begin{equation}
\mathcal{C}_{r_n}\left(\gamma_{n}\right)\triangleq\max_{0\leq \alpha \leq 1}\mathcal{C}_{r_n}\left(\gamma_{n},\alpha \right)\,,
\label{eq:chernoffalpha}\end{equation}
and the Chernoff information associated with the complete set of sensor decision functions $\underline{\gamma} \triangleq \left(\gamma_1,\ldots,\gamma_{N}\right)$ is\footnote{Note that the particular form of \eqref{eq:chernoffnet} follows due to the assumed independence of the sensor messages.}
\begin{equation}\begin{split}
\mathcal{C}\left(\,\underline{\gamma}\,\right)
& \triangleq \max_{0\leq \alpha \leq 1}\sum_{n=1}^{N}\mathcal{C}_{r_n}\left( \gamma_{n},\alpha\right)\,.
\label{eq:chernoffnet}\end{split}\end{equation}

While optimizing $P_\mathrm{E}^{(T)}=\Pr\left(\hat{H}\neq H \right)$ directly for any finite $N$ and $T$ is generally intractable, it can be shown that for the optimal FC rule it follows that \cite{Cham03}
\begin{equation}
\mathcal{C}\left(\,\underline{\gamma}\,\right)=-\lim_{T\to \infty}\frac{1}{T}\log P_\mathrm{E}^{(T)}\, .
\label{eq:exponent}\end{equation}
The Chernoff information $\mathcal{C}\left(\,\underline{\gamma}\,\right)$ at the input of the FC may thus be viewed as the (exponential) rate of which the probability of error tends to zero when increasing $T$ for a given rate allocation $\underline{r}$ and set of sensor rules $\underline{\gamma}$, and was for this reason chosen as the performance metric in \cite{Cham03}. In particular, by maximizing $\mathcal{C}\left(\,\underline{\gamma}\,\right)$ over $\underline{r}$ and $\underline{\gamma}$, Chamberland and Veeravalli \cite{Cham03} found sufficient conditions under which $N=R$ sensors each sending a one bit message (i.e., $r_1=\ldots=r_N=1$) is optimal. However, a key assumption behind their result is that the number of sensors $N$ is not fixed a priori. The problem of rate allocation for hypothesis testing remains open when the maximum number of sensors $N$ is strictly less than the rate of the MAC channel $R$. We address this question of optimal rate allocations in this regime when the method by which sensors are designed for a given rate is a-prior fixed.

To this end, we will by a \emph{sensor design method} refer to an algorithm that for any rate $r \in \mathbb{Z}^+$ generates a unique decision function which maps each input from the observation space $\mathcal{X}$ to an output from a message space $\mathcal{U}_r = \{1,2,\ldots,2^r\}$, and we will throughout the paper assume that such an algorithm exists. 
Since there given the sensor design method is a one-to-one relationship between any rate $r$ and a decision function $\gamma$, we will from now on also frequently drop the explicit mention of $\gamma_n$ in \eqref{eq:Calpha} and \eqref{eq:chernoffalpha}. In the following section we will show that if, for a given sensor design method, the resulting Chernoff information in \eqref{eq:chernoffalpha} of a single sensor, say $S_n$, is a discrete concave function of the rate $r_n$, then uniform rate allocation is an optimal rate allocation strategy.

\section{Main Results}\label{sec:res}
Suppose that there is a sensor design method which for any rate $r$ provides a decision rule with Chernoff information $\mathcal{C}_{r}$, cf.~\eqref{eq:chernoffalpha}. We will show if the resulting Chernoff information $\mathcal{C}_{r}$ is a discrete concave function of rate $r$, an optimal strategy for the design of sensors arranged as in Fig.~\ref{fig:topology} is to have sensors with the same rates $r_n=M$, where $M=R/N$ and $M,N$ and $R$ are positive integers. In what follows, we first define the concept of a discrete concave function. Then, in Theorem 1, we state our main result on the optimality of uniform rate allocation.
\begin{definition}
We say that $g : \mathbb{Z}^{+}\to \mathbb{R}$ is a discrete concave function over $\mathbb{Z}^+$ if \cite{Mur03}
\begin{equation*}g(k-1)+g(k+1)\leq 2g(k)\,, \quad \forall k\in \mathbb{Z}^{++}\,.\label{eq:concdef}\end{equation*}
\end{definition}
The following lemma follows straightforwardly for any discrete concave function $g(k)$ by iteratively using the definition above, and is given without proof.
\begin{lemma}
If $g(k)$ is a discrete concave function of $k$, then
\begin{equation}\begin{split}
g\left(k_1\right)+g\left(k_2\right)\leq g\left(\left\lceil \frac{k_1+k_2}{2}\right\rceil\right)+g\left(\left\lfloor \frac{k_1+k_2}{2}\right\rfloor\right)\,,\\
\end{split}\end{equation}
for all $k_1,k_2\in \mathbb{Z}^{+}$.
\label{lem:concavity}\end{lemma}
Lemma \ref{lem:concavity} implies that if the Chernoff information $\mathcal{C}_r$ is a discrete concave function of rate $r$, the summation of Chernoff information of two sensors with rates $r_i$ and $r_j$, is less than the summation of the Chernoff information of two sensors with rates $\left\lfloor \frac{r_i+r_j}{2}\right\rfloor$ and $ \left\lceil \frac{r_i+r_j}{2}\right\rceil$, 
where
\[\left\lceil \frac{r_i+r_j}{2}\right\rceil-\left\lfloor \frac{r_i+r_j}{2}\right\rfloor \in\{0, 1\}\,.\]
We will use this lemma for the proof of our main result in Theorem \ref{th:main}. Consider the problem of allocating rate to $N$ sensors arranged as in Fig.~\ref{fig:topology}, making iid observations about the same hypothesis $H$, and where the MAC channel is subjected to a rate constraint in \eqref{eq:Rconstraint}, where $R=MN$ for some $M\in\mathbb{Z}^{++}$. Assume that there is a (common) sensor design method which results in the Chernoff information $\mathcal{C}_{r_n}$ for sensor $S_n$ at rate $r_n$. We have the following theorem for an optimal rate allocation.
\begin{theorem}
Given a sensor design method, if for a single sensor $S_n$ the resulting Chernoff information $\mathcal{C}_{r_n}$ is a discrete concave function of rate $r_n$, a uniform rate allocation across sensors is optimal.
\label{th:main}\end{theorem}
\begin{proofs}
Consider a network of $N$ sensors with rate allocation $\underline{r} \triangleq \left(r_1,\ldots,r_N\right)$ and decision functions $\underline{\gamma} \triangleq \left(\gamma_1,\ldots,\gamma_N\right)$. Without loss of generality we can assume that $r_1\leq\ldots\leq r_N$. Let $\mathcal{C}_{r_n}$ be a discrete concave function of rate $r_n$. Consider replacing sensors $S_1$ and $S_N$, with decision functions $\gamma_1$ and $\gamma_N$, with two sensors $S'_1$ and $S'_N$ with rates $r'_1=\left\lfloor \frac{r_1+r_N}{2}\right\rfloor$ and $r'_N=\left\lceil \frac{r_1+r_N}{2}\right\rceil$, and decision functions $\gamma'_1$ and $\gamma'_N$, where $r_1\leq r'_1\leq r'_N\leq r_N$. According to Lemma \ref{lem:concavity},
\begin{equation*}
\mathcal{C}_{r_1}+\mathcal{C}_{r_N}\leq\mathcal{C}_{r'_1}+\mathcal{C}_{r'_N}\,.
\end{equation*}
By additionally letting $S'_n=S_n$, $\gamma'_n=\gamma_n$ and $r'_n=r_n$ for $n=2,\ldots,N-1$ we obtain a new rate allocation $\underline{r}'\triangleq\left(r'_1,\ldots,r'_N\right)$ and decision functions $\underline{\gamma}'\triangleq \left(\gamma'_1,\ldots,\gamma'_N\right)$ for which
\[\sum_{n=1}^N\mathcal{C}_{r_n} \leq \sum_{n=1}^N\mathcal{C}_{r'_n}\,.\]
The new rate allocation $\underline{r}'$ also satisfies the rate constraint in \eqref{eq:Rconstraint} since $r_1+r_N=r'_1+r'_N$. We can repeatedly replace the lowest-rate and the highest-rate sensors with minimum difference sensors without decreasing the Chernoff information, until we get uniform-rate sensors, i.e., $\gamma'_1=\ldots=\gamma'_N$ and $r'_1=\ldots=r'_N=M$. For the uniform rate allocation it follows that
\begin{equation}\sum_{n=1}^N \mathcal{C}_{r_n}\leq N\mathcal{C}_{M}\,.\label{eq:sumrates}\end{equation}
%
We further have that
\begin{equation}\begin{split}
\mathcal{C}\left(\,\underline{\gamma}\,\right)
&\stackrel{(a)}{\leq}  \sum_{n=1}^{N}\max_{0\leq \alpha_n \leq 1}\mathcal{C}_{r_n}\left( \alpha_n\right)\\
&\stackrel{(b)}{=} \sum_{n=1}^N \mathcal{C}_{r_n}\,
\stackrel{(c)}{\leq}  N\mathcal{C}_{M}\,,
\label{eq:proof}\end{split}\end{equation}
where $(a)$ and $(b)$ are immediate results of \eqref{eq:chernoffnet} and \eqref{eq:chernoffalpha}, respectively, and $(c)$ is obtained using \eqref{eq:sumrates}. Note however that the inequalities in \eqref{eq:proof} are satisfied with equality when all the sensors have the same rate $r_n=M$ and consequently the same decision function $\gamma = \gamma_1 = \ldots = \gamma_N$, which implies the same optimizer $\alpha^\star$ in \eqref{eq:chernoffalpha} and \eqref{eq:chernoffnet}, i.e., 
$\max_{\underline{r}}\,\mathcal{C}\left(\,\underline{\gamma} \,\right)=N\mathcal{C}_M$. \hfill \qed
\end{proofs}

As discussed above, the optimality of uniform rate allocation in a network of sensors arranged as in Fig.~\ref{fig:topology} relies on the concavity of the Chernoff information of the sensor design method. If there is no such a design method, the results of this paper are in vain. We will therefore explore this point numerically in the next section under the assumption of equal variance Gaussian observations.


\section{Numerical Results}\label{sec:sim}
We shall first consider the design method of Benitz and Bucklew \cite{Ben89} for the design of sensor decisions, and numerically show that the Chernoff information resulting from their method is a discrete concave function of rate. Moreover, using a numerical optimization method we design sensor decision functions with good performance, and show that the concavity remains. Finally, using simulations we relate this to the error probability performance of different rate allocations in a network of sensors.

We consider the case where each observation $x_{n,t}$ consists of an antipodal signal $\pm m$ in an additive unit-variance white Gaussian noise $v_n$. The observation model is
\begin{equation}
\begin{split}
&H_0: {x_n}={-m} +{v_n}\,, \\
&H_1: {x_n}={+m} + {v_n}\,.
\label{eq:observmodel}\end{split}
\end{equation}
The observation space at each sensor is in this case equal to the real space, i.e., $\mathcal{X}=\mathbb{R}$.
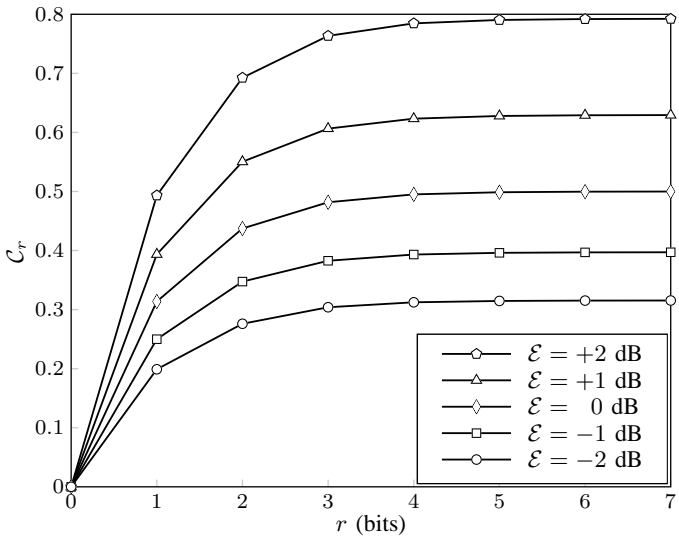
\begin{figure}[t]
\centering
%
%
%
\psset{xunit=0.142857\plotwidth,yunit=0.985887\plotwidth}%
\begin{pspicture}(-0.774194,-0.088889)(7.080645,0.818713)%


\psline[linewidth=\AxesLineWidth,linecolor=GridColor](0.000000,0.000000)(0.000000,0.012172)
\psline[linewidth=\AxesLineWidth,linecolor=GridColor](1.000000,0.000000)(1.000000,0.012172)
\psline[linewidth=\AxesLineWidth,linecolor=GridColor](2.000000,0.000000)(2.000000,0.012172)
\psline[linewidth=\AxesLineWidth,linecolor=GridColor](3.000000,0.000000)(3.000000,0.012172)
\psline[linewidth=\AxesLineWidth,linecolor=GridColor](4.000000,0.000000)(4.000000,0.012172)
\psline[linewidth=\AxesLineWidth,linecolor=GridColor](5.000000,0.000000)(5.000000,0.012172)
\psline[linewidth=\AxesLineWidth,linecolor=GridColor](6.000000,0.000000)(6.000000,0.012172)
\psline[linewidth=\AxesLineWidth,linecolor=GridColor](7.000000,0.000000)(7.000000,0.012172)
\psline[linewidth=\AxesLineWidth,linecolor=GridColor](0.000000,0.000000)(0.084000,0.000000)
\psline[linewidth=\AxesLineWidth,linecolor=GridColor](0.000000,0.100000)(0.084000,0.100000)
\psline[linewidth=\AxesLineWidth,linecolor=GridColor](0.000000,0.200000)(0.084000,0.200000)
\psline[linewidth=\AxesLineWidth,linecolor=GridColor](0.000000,0.300000)(0.084000,0.300000)
\psline[linewidth=\AxesLineWidth,linecolor=GridColor](0.000000,0.400000)(0.084000,0.400000)
\psline[linewidth=\AxesLineWidth,linecolor=GridColor](0.000000,0.500000)(0.084000,0.500000)
\psline[linewidth=\AxesLineWidth,linecolor=GridColor](0.000000,0.600000)(0.084000,0.600000)
\psline[linewidth=\AxesLineWidth,linecolor=GridColor](0.000000,0.700000)(0.084000,0.700000)
\psline[linewidth=\AxesLineWidth,linecolor=GridColor](0.000000,0.800000)(0.084000,0.800000)

{ \footnotesize 
\rput[t](0.000000,-0.012172){$0$}
\rput[t](1.000000,-0.012172){$1$}
\rput[t](2.000000,-0.012172){$2$}
\rput[t](3.000000,-0.012172){$3$}
\rput[t](4.000000,-0.012172){$4$}
\rput[t](5.000000,-0.012172){$5$}
\rput[t](6.000000,-0.012172){$6$}
\rput[t](7.000000,-0.012172){$7$}
\rput[r](-0.084000,0.000000){$0$}
\rput[r](-0.084000,0.100000){$0.1$}
\rput[r](-0.084000,0.200000){$0.2$}
\rput[r](-0.084000,0.300000){$0.3$}
\rput[r](-0.084000,0.400000){$0.4$}
\rput[r](-0.084000,0.500000){$0.5$}
\rput[r](-0.084000,0.600000){$0.6$}
\rput[r](-0.084000,0.700000){$0.7$}
\rput[r](-0.084000,0.800000){$0.8$}
} 

\psframe[linewidth=\AxesLineWidth,dimen=middle](0.000000,0.000000)(7.000000,0.800000)

{ \small 
\rput[b](3.500000,-0.088889){
\begin{tabular}{c}
$r$ (bits)\\
\end{tabular}
}

\rput[t]{90}(-0.774194,0.400000){
\begin{tabular}{c}
$\mathcal{C}_r$\\
\end{tabular}
}
} 

\newrgbcolor{color267.0032}{0  0  0}
\psline[plotstyle=line,linejoin=1,showpoints=true,dotstyle=Bpentagon,dotsize=\MarkerSize,linestyle=solid,linewidth=\LineWidth,linecolor=color267.0032]
(0.000000,0.000000)(1.000000,0.493321)(2.000000,0.692597)(3.000000,0.763466)(4.000000,0.784642)
(5.000000,0.790421)(6.000000,0.791931)(7.000000,0.792316)

\newrgbcolor{color268.0027}{0  0  0}
\psline[plotstyle=line,linejoin=1,showpoints=true,dotstyle=Btriangle,dotsize=\MarkerSize,linestyle=solid,linewidth=\LineWidth,linecolor=color268.0027]
(0.000000,0.000000)(1.000000,0.393574)(2.000000,0.550382)(3.000000,0.606480)(4.000000,0.623269)
(5.000000,0.627855)(6.000000,0.629053)(7.000000,0.629359)

\newrgbcolor{color269.0028}{0  0  0}
\psline[plotstyle=line,linejoin=1,showpoints=true,dotstyle=diamond,dotsize=\MarkerSize,linestyle=solid,linewidth=\LineWidth,linecolor=color269.0028]
(0.000000,0.000000)(1.000000,0.313741)(2.000000,0.437325)(3.000000,0.481768)(4.000000,0.495084)
(5.000000,0.498723)(6.000000,0.499675)(7.000000,0.499918)

\newrgbcolor{color270.0028}{0  0  0}
\psline[plotstyle=line,linejoin=1,showpoints=true,dotstyle=Bsquare,dotsize=\MarkerSize,linestyle=solid,linewidth=\LineWidth,linecolor=color270.0028]
(0.000000,0.000000)(1.000000,0.249932)(2.000000,0.347465)(3.000000,0.382696)(4.000000,0.393261)
(5.000000,0.396150)(6.000000,0.396906)(7.000000,0.397099)

\newrgbcolor{color271.0028}{0  0  0}
\psline[plotstyle=line,linejoin=1,showpoints=true,dotstyle=Bo,dotsize=\MarkerSize,linestyle=solid,linewidth=\LineWidth,linecolor=color271.0028]
(0.000000,0.000000)(1.000000,0.198990)(2.000000,0.276054)(3.000000,0.303995)(4.000000,0.312380)
(5.000000,0.314673)(6.000000,0.315273)(7.000000,0.315427)

{ \small 
\rput(5.5,0.131508){%
\psshadowbox[framesep=0pt,shadowsize=0pt,linewidth=\AxesLineWidth]{\psframebox*{\begin{tabular}{l}
\Rnode{a1}{\hspace*{0.0ex}} \hspace*{0.7cm} \Rnode{a2} {~~$ \mathcal{E}= +2$ dB} \\
\Rnode{a3}{\hspace*{0.0ex}} \hspace*{0.7cm} \Rnode{a4} {~~$ \mathcal{E}= +1$ dB} \\
\Rnode{a5}{\hspace*{0.0ex}} \hspace*{0.7cm} \Rnode{a6} {~~$ \mathcal{E}=\quad\!\! 0$ dB} \\
\Rnode{a7}{\hspace*{0.0ex}} \hspace*{0.7cm} \Rnode{a8} {~~$ \mathcal{E}= -1$ dB} \\
\Rnode{a9}{\hspace*{0.0ex}} \hspace*{0.7cm} \Rnode{a10}{~~$ \mathcal{E}= -2$ dB} \\
\end{tabular}}
\ncline[linestyle=solid,linewidth=\LineWidth,linecolor=color267.0032]{a1}{a2} \ncput{\psdot[dotstyle=Bpentagon,dotsize=\MarkerSize,linecolor=color267.0032]}
\ncline[linestyle=solid,linewidth=\LineWidth,linecolor=color268.0027]{a3}{a4} \ncput{\psdot[dotstyle=Btriangle,dotsize=\MarkerSize,linecolor=color268.0027]}
\ncline[linestyle=solid,linewidth=\LineWidth,linecolor=color269.0028]{a5}{a6} \ncput{\psdot[dotstyle=diamond,dotsize=\MarkerSize,linecolor=color269.0028]}
\ncline[linestyle=solid,linewidth=\LineWidth,linecolor=color270.0028]{a7}{a8} \ncput{\psdot[dotstyle=Bsquare,dotsize=\MarkerSize,linecolor=color270.0028]}
\ncline[linestyle=solid,linewidth=\LineWidth,linecolor=color271.0028]{a9}{a10} \ncput{\psdot[dotstyle=Bo,dotsize=\MarkerSize,linecolor=color271.0028]}
}%
}%
} 

\end{pspicture}%

\caption{Chernoff information of a single sensor designed using Benitz and Bucklew's method as a function of rate, for different SNRs.}
\label{fig:BBconcavity}\end{figure}

\subsection{Benitz and Bucklew's Method}
Benitz and Bucklew \cite{Ben89} proposed a sensor design method (or quantization rule) in detection with iid observations, using a companding function $q : \mathcal{X} \to [0,1]$. The idea behind the method is to uniformly quantize the range $[0,1]$, and let the companding function define the quantization of $\mathcal{X}$. The optimal companding function $q$ depends on the conditional distributions of the observations, see \cite[Section V]{Ben89}, and the key result of \cite{Ben89} is a set of conditions that identify the asymptotically optimal $q$ in terms of Chernoff information in the high rate regime where $r \rightarrow \infty$, but the design methods have been empirically observed to work well also for finite $r$.


For the observation model in \eqref{eq:observmodel} it can, following the general derivation of \cite{Ben89}, be shown that the asymptotically optimal companding function $q$ is given by
$$
q(x) = \mathcal{G} \left( \frac{x}{\sqrt{3}} \right) \, ,
$$
where $\mathcal{G}(y)$ is the unit-variance Gaussian cumulative distribution function given by
$$
\mathcal{G}(y)=\int_{-\infty}^y \!\frac{1}{\sqrt{2\pi}}e^{-\frac{\,\,t^2}{2\,}}\,dt\,.
$$
This result holds, somewhat surprisingly, independently of the specific value of $m$. By the monotonicity of $q(x)$ it follows that the obtained quantizer of $\mathcal{X} = \mathbb{R}$ is a monotone quantizer \cite{Tsi93Ext} with
partitions $\mathcal{I}_1\triangleq[-\infty,b_1),\mathcal{I}_2\triangleq[b_1,b_2),\ldots,\mathcal{I}_K\triangleq[b_{K-1},\infty]$ for $K \triangleq \mathrm 2^r$,
with boundaries given by $b_i=\sqrt{3}\mathcal{G}^{-1}\left( i/K \right)$ for $i=1,\ldots,K-1$. The resulting Chernoff information of a rate-$r$ sensor becomes $\mathcal{C}_r = \mathcal{C}_r^\star + o(2^{-2r})$ where \cite{Ben89}
\begin{equation}
\mathcal{C}_r^\star =\frac{m^2}{2}-\log\left[1+\frac{\pi\sqrt{3}\,m^2}{4}\,2^{-2r} \right]\,.
\label{eq:BenChernoff}\end{equation}
At high rates, the term $o(2^{-2r})$ vanishes, and it can be shown from first principles that
\[\mathcal{C}_{r-1}^\star+\mathcal{C}_{r+1}^\star \leq 2\mathcal{C}_r ^\star\,,\]
i.e., the Chernoff information of sensors designed using this method is asymptotically a discrete concave function of rate $r$. Although it is difficult to formally prove concavity for finite $r$, it is straightforward to calculate $\mathcal{C}_{r_n}$ in \eqref{eq:chernoffalpha} for any given $r$ and $m$.
Fig.~\ref{fig:BBconcavity} shows the Chernoff information of a sensor designed using the method of \cite{Ben89} as a function of rate, for different values of the per channel signal-to-noise ratio (SNR), which we define as $\mathcal{E}\triangleq \vert m \vert ^2$, and provides empirical evidence of concavity. The Chernoff information $\mathcal{C}_r$ is for any rate $r$ upper bounded by the Chernoff information $\mathcal{C}_\infty$ contained in the raw observation, which is defined as
\begin{equation}
\label{eq:Cinf}
\mathcal{C}_\infty\triangleq \max_{0\leq \alpha \leq 1} -\int_\mathcal{X}{f_{X\vert H}(x\vert H_0)}^\alpha {f_{X\vert H}(x\vert H_1)}^{1-\alpha}\,dx\, .
\end{equation}
For Gaussian distributed observations $\mathcal{C}_\infty=m^2/2$, which can also be obtained from \eqref{eq:BenChernoff} by letting $r\to \infty$.

\subsection{Numerical Method}

In order to provide a contrast to the previous section, we also consider a quantizer obtained through a numerical optimization. For a rate-$r$ quantizer,  we form a partition of the real interval into $K=2^r$ intervals $\mathcal{I}_1,\ldots,\mathcal{I}_K$ with randomly\footnote{Boundaries are drawn uniformly in the range $[-m-5,m+5]$ and sorted.} generated boundaries $\{b_1,\ldots,b_{K-1}\}$. Assume that $b_0\triangleq-\infty \leq b_1\leq\ldots\leq b_{K-1}\leq b_{K}\triangleq+\infty$. We then iteratively update the boundaries and the value of $\alpha$ in \eqref{eq:Calpha} in such a way that $C_{r_n}(\gamma_n,\alpha)$ is maximized at each step. A full iteration consists of updating the boundaries and updating the value of $\alpha$. In each iteration, the values of each $b_i$ -- from $b_1$ to $b_{K-1}$ -- are first updated sequentially while the other boundaries and $\alpha$ are kept fixed. The value of $\alpha$ is then updated while the boundaries are kept fixed. The position of a boundary, say $b_i$, is modified in the interval $[b_{i-1},b_{i+1}]$ to (numerically) maximize $C_{r_n}(\gamma_n,\alpha)$, and the value of $\alpha$ is selected to (numerically) maximize $C_{r_n}(\gamma_n,\alpha)$ over $\alpha \in [0,1]$. The iterations are continued until the improvement in $C_{r_n}(\gamma_n,\alpha)$ is less than $\eta = 10^{-4}$.

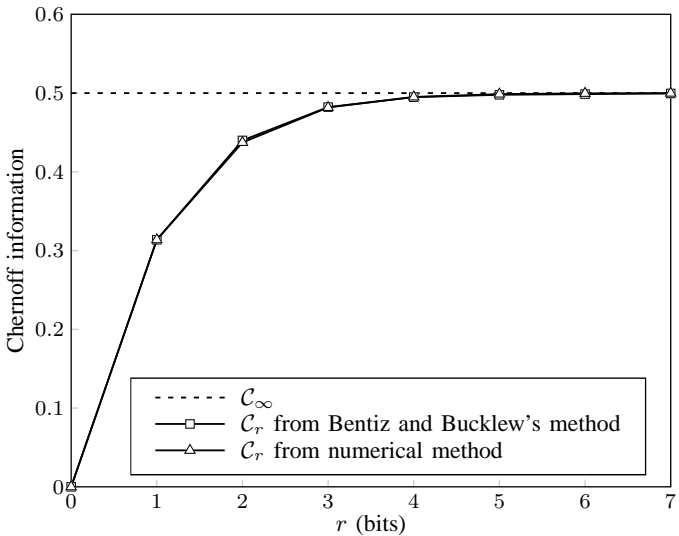
\begin{figure}[t]
\centering
%
%
\psset{xunit=0.142857\plotwidth,yunit=1.314297\plotwidth}%
\begin{pspicture}(-0.774194,-0.066678)(7.080645,0.614137)%


\psline[linewidth=\AxesLineWidth,linecolor=GridColor](0.000000,0.000000)(0.000000,0.009130)
\psline[linewidth=\AxesLineWidth,linecolor=GridColor](1.000000,0.000000)(1.000000,0.009130)
\psline[linewidth=\AxesLineWidth,linecolor=GridColor](2.000000,0.000000)(2.000000,0.009130)
\psline[linewidth=\AxesLineWidth,linecolor=GridColor](3.000000,0.000000)(3.000000,0.009130)
\psline[linewidth=\AxesLineWidth,linecolor=GridColor](4.000000,0.000000)(4.000000,0.009130)
\psline[linewidth=\AxesLineWidth,linecolor=GridColor](5.000000,0.000000)(5.000000,0.009130)
\psline[linewidth=\AxesLineWidth,linecolor=GridColor](6.000000,0.000000)(6.000000,0.009130)
\psline[linewidth=\AxesLineWidth,linecolor=GridColor](7.000000,0.000000)(7.000000,0.009130)
\psline[linewidth=\AxesLineWidth,linecolor=GridColor](0.000000,0.000000)(0.084000,0.000000)
\psline[linewidth=\AxesLineWidth,linecolor=GridColor](0.000000,0.100000)(0.084000,0.100000)
\psline[linewidth=\AxesLineWidth,linecolor=GridColor](0.000000,0.200000)(0.084000,0.200000)
\psline[linewidth=\AxesLineWidth,linecolor=GridColor](0.000000,0.300000)(0.084000,0.300000)
\psline[linewidth=\AxesLineWidth,linecolor=GridColor](0.000000,0.400000)(0.084000,0.400000)
\psline[linewidth=\AxesLineWidth,linecolor=GridColor](0.000000,0.500000)(0.084000,0.500000)
\psline[linewidth=\AxesLineWidth,linecolor=GridColor](0.000000,0.600000)(0.084000,0.600000)

{ \footnotesize 
\rput[t](0.000000,-0.009130){$0$}
\rput[t](1.000000,-0.009130){$1$}
\rput[t](2.000000,-0.009130){$2$}
\rput[t](3.000000,-0.009130){$3$}
\rput[t](4.000000,-0.009130){$4$}
\rput[t](5.000000,-0.009130){$5$}
\rput[t](6.000000,-0.009130){$6$}
\rput[t](7.000000,-0.009130){$7$}
\rput[r](-0.084000,0.000000){$0$}
\rput[r](-0.084000,0.100000){$0.1$}
\rput[r](-0.084000,0.200000){$0.2$}
\rput[r](-0.084000,0.300000){$0.3$}
\rput[r](-0.084000,0.400000){$0.4$}
\rput[r](-0.084000,0.500000){$0.5$}
\rput[r](-0.084000,0.600000){$0.6$}
} 

\psframe[linewidth=\AxesLineWidth,dimen=middle](0.000000,0.000000)(7.000000,0.600100)

{ \small 
\rput[b](3.500000,-0.066678){
\begin{tabular}{c}
$r$ (bits)\\
\end{tabular}
}

\rput[t]{90}(-0.774194,0.300050){
\begin{tabular}{c}
Chernoff information\\
\end{tabular}
}
} 

\newrgbcolor{color314.0032}{0  0  0}
\psline[plotstyle=line,linejoin=1,linestyle=dashed,dash=2pt 3pt,linewidth=\LineWidth,linecolor=color314.0032]
(0.000000,0.500000)(7.000000,0.500000)

\newrgbcolor{color315.0027}{0  0  0}
\psline[plotstyle=line,linejoin=1,showpoints=true,dotstyle=Bsquare,dotsize=\MarkerSize,linestyle=solid,linewidth=\LineWidth,linecolor=color315.0027]
(0.000000,0.000000)(1.000000,0.313741)(2.000000,0.439926)(3.000000,0.482240)(4.000000,0.494887)
(5.000000,0.497850)(6.000000,0.498752)(7.000000,0.499549)

\newrgbcolor{color316.0027}{0  0  0}
\psline[plotstyle=line,linejoin=1,showpoints=true,dotstyle=Btriangle,dotsize=\MarkerSize,linestyle=solid,linewidth=\LineWidth,linecolor=color316.0027]
(0.000000,0.000000)(1.000000,0.313741)(2.000000,0.437325)(3.000000,0.481768)(4.000000,0.495084)
(5.000000,0.498723)(6.000000,0.499675)(7.000000,0.499918)

{ \small 
\rput(3.7,0.076788){%
\psshadowbox[framesep=0pt,shadowsize=0pt,linewidth=\AxesLineWidth]{\psframebox*{\begin{tabular}{l}
\Rnode{a1}{\hspace*{0.0ex}} \hspace*{0.7cm} \Rnode{a2}{~~$\mathcal{C}_\infty$} \\
\Rnode{a3}{\hspace*{0.0ex}} \hspace*{0.7cm} \Rnode{a4}{~~$\mathcal{C}_r$ from Bentiz and Bucklew's method} \\
\Rnode{a5}{\hspace*{0.0ex}} \hspace*{0.7cm} \Rnode{a6}{~~$\mathcal{C}_r$ from numerical method} \\
\end{tabular}}
\ncline[linestyle=dashed,dash=2pt 3pt,linewidth=\LineWidth,linecolor=color314.0032]{a1}{a2}
\ncline[linestyle=solid,linewidth=\LineWidth,linecolor=color315.0027]{a3}{a4} \ncput{\psdot[dotstyle=Bsquare,dotsize=\MarkerSize,linecolor=color315.0027]}
\ncline[linestyle=solid,linewidth=\LineWidth,linecolor=color316.0027]{a5}{a6} \ncput{\psdot[dotstyle=Btriangle,dotsize=\MarkerSize,linecolor=color316.0027]}
}%
}%
} 

\end{pspicture}%

\caption{Chernoff information of a single sensor designed using Benitz and Bucklew's method and using the numerical method, and the Chernoff information contained in each observation, for $\mathcal{E}=0$ dB.}
\label{fig:Sim}\end{figure}

Fig.~\ref{fig:Sim} illustrates the Chernoff information contained in an observation [cf.~\eqref{eq:Cinf}], the Chernoff information of a sensor designed using the numerical method described above, and the Chernoff information of a sensor designed using the method proposed in \cite{Ben89}, when $m=1$ (or $\mathcal{E}=0$ dB). The numerical method always results in $\alpha^\star=0.5$ for any rate $r$, which is implicitly used also in \eqref{eq:BenChernoff} given the optimality $\alpha = 0.5$ in \eqref{eq:Cinf}. Two things can be observed in Fig.~\ref{fig:Sim}: The resulting Chernoff information of the numerically designed sensors is discrete concave; and the difference to the asymptotic method of Benitz and Bucklew is marginal. We could naturally also initialize the numerical optimization with the result of \cite{Ben89}, but the difference by doing so is, again, marginal.





\subsection{The Error Probability Performance of Sensor Networks}

Finally, in order to illustrate the usefulness of the obtained results for a finite $T$, we consider the probability of error of a network of sensors designed using the numerical design method described in the previous section, for the case of a single shot observation, i.e., $T=1$.  We explicitly consider a network of $N=6$ sensors arranged as in Fig.~\ref{fig:topology} with the same observation model as before, and study the effect of different rate allocations $\underline{r} = (r_1,\ldots,r_N)$ subject to the rate constraint in \eqref{eq:Rconstraint} with $R=12$ bits. The decision functions are redesigned for each SNR value $\mathcal{E} = |m|^2$. 
It is worth noting that as the numerical method yields $\alpha^\star = 0.5$ for all rates $r$, the inequalities in \eqref{eq:proof} will still be tight meaning that the Chernoff information of the whole network is in this case given by the sum of the Chernoff information of the individual sensors, although this can not be assumed in general for networks with non-uniform rate allocations.



The probability of error $P_\mathrm{E} = P_\mathrm{E}^{(1)}$ of the MAP FC rule can for a given set of decision functions, and under the assumption of equally likely hypotheses, be obtained as (cf. \cite{Alla14})
\begin{equation*}
P_\mathrm{E}=1-\frac{1}{2}\sum_{\underline{u}} \max_{j=0,1}\left\{P_{\underline{U}\vert H}\left( \underline{u}\vert H_j\right)\right\} ,
\end{equation*}
which can be straightforwardly computed numerically for the examples at hand using the Gaussian $\mathcal{Q}$-function and the obtained quantization thresholds, without the need for Monte-Carlo simulations.


\begin{figure}[t]
\centering
%
%
\psset{xunit=0.100000\plotwidth,yunit=0.175269\plotwidth}%
\begin{pspicture}(-6.198157,-5.500000)(5.115207,-0.394737)%


\psline[linewidth=\AxesLineWidth,linecolor=GridColor](-5.000000,-5.000000)(-5.000000,-4.931534)
\psline[linewidth=\AxesLineWidth,linecolor=GridColor](-4.000000,-5.000000)(-4.000000,-4.931534)
\psline[linewidth=\AxesLineWidth,linecolor=GridColor](-3.000000,-5.000000)(-3.000000,-4.931534)
\psline[linewidth=\AxesLineWidth,linecolor=GridColor](-2.000000,-5.000000)(-2.000000,-4.931534)
\psline[linewidth=\AxesLineWidth,linecolor=GridColor](-1.000000,-5.000000)(-1.000000,-4.931534)
\psline[linewidth=\AxesLineWidth,linecolor=GridColor](0.000000,-5.000000)(0.000000,-4.931534)
\psline[linewidth=\AxesLineWidth,linecolor=GridColor](1.000000,-5.000000)(1.000000,-4.931534)
\psline[linewidth=\AxesLineWidth,linecolor=GridColor](2.000000,-5.000000)(2.000000,-4.931534)
\psline[linewidth=\AxesLineWidth,linecolor=GridColor](3.000000,-5.000000)(3.000000,-4.931534)
\psline[linewidth=\AxesLineWidth,linecolor=GridColor](4.000000,-5.000000)(4.000000,-4.931534)
\psline[linewidth=\AxesLineWidth,linecolor=GridColor](5.000000,-5.000000)(5.000000,-4.931534)
\psline[linewidth=\AxesLineWidth,linecolor=GridColor](-5.000000,-5.000000)(-4.880000,-5.000000)
\psline[linewidth=\AxesLineWidth,linecolor=GridColor](-5.000000,-4.500000)(-4.880000,-4.500000)
\psline[linewidth=\AxesLineWidth,linecolor=GridColor](-5.000000,-4.000000)(-4.880000,-4.000000)
\psline[linewidth=\AxesLineWidth,linecolor=GridColor](-5.000000,-3.500000)(-4.880000,-3.500000)
\psline[linewidth=\AxesLineWidth,linecolor=GridColor](-5.000000,-3.000000)(-4.880000,-3.000000)
\psline[linewidth=\AxesLineWidth,linecolor=GridColor](-5.000000,-2.500000)(-4.880000,-2.500000)
\psline[linewidth=\AxesLineWidth,linecolor=GridColor](-5.000000,-2.000000)(-4.880000,-2.000000)
\psline[linewidth=\AxesLineWidth,linecolor=GridColor](-5.000000,-1.500000)(-4.880000,-1.500000)
\psline[linewidth=\AxesLineWidth,linecolor=GridColor](-5.000000,-1.000000)(-4.880000,-1.000000)
\psline[linewidth=\AxesLineWidth,linecolor=GridColor](-5.000000,-0.500000)(-4.880000,-0.500000)

{ \footnotesize 
\rput[t](-5.000000,-5.068466){$-5$}
\rput[t](-4.000000,-5.068466){$-4$}
\rput[t](-3.000000,-5.068466){$-3$}
\rput[t](-2.000000,-5.068466){$-2$}
\rput[t](-1.000000,-5.068466){$-1$}
\rput[t](0.000000,-5.068466){$0$}
\rput[t](1.000000,-5.068466){$1$}
\rput[t](2.000000,-5.068466){$2$}
\rput[t](3.000000,-5.068466){$3$}
\rput[t](4.000000,-5.068466){$4$}
\rput[t](5.000000,-5.068466){$5$}
\rput[r](-5.120000,-5.000000){$-5$}
\rput[r](-5.120000,-4.500000){$-4.5$}
\rput[r](-5.120000,-4.000000){$-4$}
\rput[r](-5.120000,-3.500000){$-3.5$}
\rput[r](-5.120000,-3.000000){$-3$}
\rput[r](-5.120000,-2.500000){$-2.5$}
\rput[r](-5.120000,-2.000000){$-2$}
\rput[r](-5.120000,-1.500000){$-1.5$}
\rput[r](-5.120000,-1.000000){$-1$}
\rput[r](-5.120000,-0.500000){$-0.5$}
} 

\psframe[linewidth=\AxesLineWidth,dimen=middle](-5.000000,-5.000000)(5.000000,-0.500000)

{ \small 
\rput[b](0.000000,-5.500000){
\begin{tabular}{c}
$\mathcal{E}$ (dB)\\
\end{tabular}
}

\rput[t]{90}(-6.198157,-2.750000){
\begin{tabular}{c}
$\log_{10} P_{\mathrm{E}}$\\
\end{tabular}
}
} 

\newrgbcolor{color1.0027}{0  0  0}
\psline[plotstyle=line,linejoin=1,showpoints=true,dotstyle=diamond,dotsize=\MarkerSize,linestyle=solid,linewidth=\LineWidth,linecolor=color1.0027]
(-5.000000,-0.780510)(-4.000000,-0.860243)(-3.000000,-0.955596)(-2.000000,-1.070125)(-1.000000,-1.208289)
(0.000000,-1.375679)(1.000000,-1.579323)(2.000000,-1.828065)(3.000000,-2.133041)(4.000000,-2.508278)
(5.000000,-2.971459)

\newrgbcolor{color2.0022}{0  0  0}
\psline[plotstyle=line,linejoin=1,showpoints=true,dotstyle=Bpentagon,dotsize=\MarkerSize,linestyle=solid,linewidth=\LineWidth,linecolor=color2.0022]
(-5.000000,-0.870257)(-4.000000,-0.967597)(-3.000000,-1.084570)(-2.000000,-1.225749)(-1.000000,-1.396870)
(0.000000,-1.605144)(1.000000,-1.859640)(2.000000,-2.171777)(3.000000,-2.555933)(4.000000,-3.030213)
(5.000000,-3.617399)

\newrgbcolor{color3.0022}{0  0  0}
\psline[plotstyle=line,linejoin=1,showpoints=true,dotstyle=Btriangle,dotsize=\MarkerSize,linestyle=solid,linewidth=\LineWidth,linecolor=color3.0022]
(-5.000000,-0.940064)(-4.000000,-1.051071)(-3.000000,-1.184708)(-2.000000,-1.346263)(-1.000000,-1.542300)
(0.000000,-1.780948)(1.000000,-2.072369)(2.000000,-2.429131)(3.000000,-2.866738)(4.000000,-3.405440)
(5.000000,-4.068954)

\newrgbcolor{color4.0022}{0  0  0}
\psline[plotstyle=line,linejoin=1,showpoints=true,dotstyle=Bsquare,dotsize=\MarkerSize,linestyle=solid,linewidth=\LineWidth,linecolor=color4.0022]
(-5.000000,-0.981888)(-4.000000,-1.101627)(-3.000000,-1.246139)(-2.000000,-1.421262)(-1.000000,-1.634308)
(0.000000,-1.894448)(1.000000,-2.213117)(2.000000,-2.604712)(3.000000,-3.087053)(4.000000,-3.682216)
(5.000000,-4.418104)

\newrgbcolor{color5.0022}{0  0  0}
\psline[plotstyle=line,linejoin=1,showpoints=true,dotstyle=Bo,dotsize=\MarkerSize,linestyle=solid,linewidth=\LineWidth,linecolor=color5.0022]
(-5.000000,-1.003099)(-4.000000,-1.127313)(-3.000000,-1.277415)(-2.000000,-1.459559)(-1.000000,-1.681469)
(0.000000,-1.952850)(1.000000,-2.285884)(2.000000,-2.695851)(3.000000,-3.201877)(4.000000,-3.827819)
(5.000000,-4.603266)

{ \small 
\rput(-2.3,-4.182361){%
\psshadowbox[framesep=0pt,shadowsize=0pt,linewidth=\AxesLineWidth]{\psframebox*{\begin{tabular}{l}
\Rnode{a1}{\hspace*{0.0ex}} \hspace*{0.7cm} \Rnode{a2}{~~$[4,\,  4,\,  4,\,  0,\, 0,\,  0]$} \\
\Rnode{a3}{\hspace*{0.0ex}} \hspace*{0.7cm} \Rnode{a4}{~~$[3,\,  3,\,  3,\,  3,\,  0,\,  0]$} \\
\Rnode{a5}{\hspace*{0.0ex}} \hspace*{0.7cm} \Rnode{a6}{~~$[5,\,  3,\,  1,\,  1,\,  1,\,  1]$} \\
\Rnode{a7}{\hspace*{0.0ex}} \hspace*{0.7cm} \Rnode{a8}{~~$[3,\,  3,\,  2,\,  2,\,  1,\,  1]$} \\
\Rnode{a9}{\hspace*{0.0ex}} \hspace*{0.7cm} \Rnode{a10}{~~$[2,\,  2,\,  2,\,  2,\,  2,\,  2]$} \\
\end{tabular}}
\ncline[linestyle=solid,linewidth=\LineWidth,linecolor=color1.0027]{a1}{a2} \ncput{\psdot[dotstyle=diamond,dotsize=\MarkerSize,linecolor=color1.0027]}
\ncline[linestyle=solid,linewidth=\LineWidth,linecolor=color2.0022]{a3}{a4} \ncput{\psdot[dotstyle=Bpentagon,dotsize=\MarkerSize,linecolor=color2.0022]}
\ncline[linestyle=solid,linewidth=\LineWidth,linecolor=color3.0022]{a5}{a6} \ncput{\psdot[dotstyle=Btriangle,dotsize=\MarkerSize,linecolor=color3.0022]}
\ncline[linestyle=solid,linewidth=\LineWidth,linecolor=color4.0022]{a7}{a8} \ncput{\psdot[dotstyle=Bsquare,dotsize=\MarkerSize,linecolor=color4.0022]}
\ncline[linestyle=solid,linewidth=\LineWidth,linecolor=color5.0022]{a9}{a10} \ncput{\psdot[dotstyle=Bo,dotsize=\MarkerSize,linecolor=color5.0022]}
}%
}%
} 

\end{pspicture}%

\caption{Error probability performance of a sensor network for different rate allocation schemes.}
\label{fig:Pe}\end{figure}
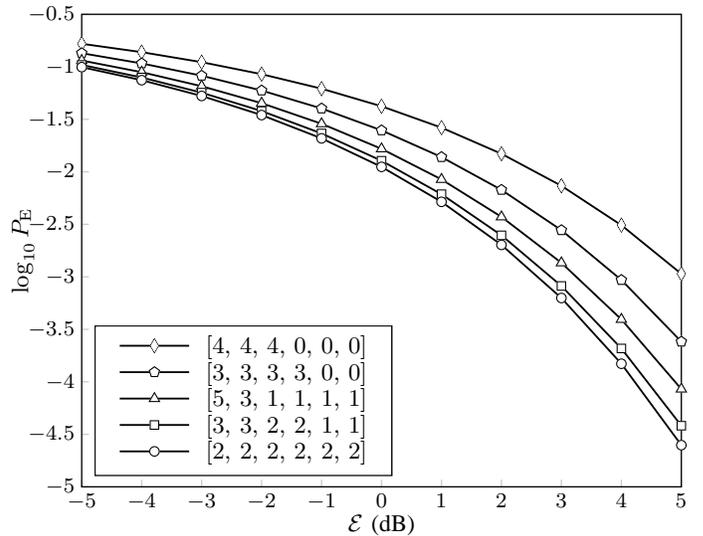

The resulting probability of error for different rate allocations are illustrated in Fig.~\ref{fig:Pe}. It can be observed that the uniform rate allocation outperforms all the other rate allocation schemes in terms of the error probability performance, which is consistent with the results obtained by studying the Chernoff information. 

\section{Conclusion} \label{sec:con}

We have in this paper obtained a sufficient condition for the optimality of uniform rate allocations for sum-rate constrained decentralized detection in wireless sensor networks, and then numerically verified that this condition holds true for some example sensor design methods.  Although it is in general hard to stringently prove the required concavity property, we have in \cite{Alla15TSP} obtained simplified sufficient condition for the discrete concavity of the Bhattacharyya distance, obtained from \eqref{eq:Calpha} with $\alpha = 0.5$, completed the stringent proof under a Laplacian observation model and under the Gaussian model given the truth of a conjecture regarding the Gaussian $\mathcal{Q}$-function.
%
\section*{Acknowledgment}
This work has been supported in part by the ACCESS seed project DeWiNe.\vspace{-.05em}

\bibliographystyle{IEEEtran}
\bibliography{Ref}

\end{document}